\documentstyle[12pt]{article}
\def\Msun{\ifmmode{~M_\odot}\else$M_\odot$~\fi}
\def\kms{\ifmmode{$~km\thinspace s$^{-1}~}\else km\thinspace s$^{-1}~$\fi}
\def\ga{\mathrel{\mathchoice {\vcenter{\offinterlineskip\halign{\hfil
$\displaystyle##$\hfil\cr>\cr\noalign{\vskip1.5pt}\sim\cr}}}
{\vcenter{\offinterlineskip\halign{\hfil$\textstyle##$\hfil\cr>\cr
\noalign{\vskip1.0pt}\sim\cr}}}
{\vcenter{\offinterlineskip\halign{\hfil$\scriptstyle##$\hfil\cr>\cr
\noalign{\vskip0.5pt}\sim\cr}}}
{\vcenter{\offinterlineskip\halign{\hfil$\scriptscriptstyle##$\hfil
\cr>\cr\noalign{\vskip0.5pt}\sim\cr}}}}}
\def\la{\mathrel{\mathchoice {\vcenter{\offinterlineskip\halign{\hfil
$\displaystyle##$\hfil\cr<\cr\noalign{\vskip1.5pt}\sim\cr}}}
{\vcenter{\offinterlineskip\halign{\hfil$\textstyle##$\hfil\cr<\cr
\noalign{\vskip1.0pt}\sim\cr}}}
{\vcenter{\offinterlineskip\halign{\hfil$\scriptstyle##$\hfil\cr<\cr
\noalign{\vskip0.5pt}\sim\cr}}}
{\vcenter{\offinterlineskip\halign{\hfil$\scriptscriptstyle##$\hfil
\cr<\cr\noalign{\vskip0.5pt}\sim\cr}}}}}
\def\appendix{\par
 \setcounter{section}{0}
 \def\thesection{A\arabic{section}}
 \setcounter{equation}{0}
 \def\theequation{A\arabic{equation}}
 \setcounter{figure}{0}
 \def\thefigure{A\@arabic\c@figure}
 \setcounter{table}{0}
 \def\thetable{A\@arabic\c@table}
}

\begin{document}

\title{{\bf The structure of isothermal, self-gravitating, stationary gas
spheres for
softened gravity}}
\author{\\
Jesper Sommer-Larsen$^{1}$,
Henrik Vedel$^{1}$ and
\and Uffe Hellsten$^{2}$\\
\\
$^1$Theoretical Astrophysics Center\\
Juliane Maries Vej 30, DK-2100 Copenhagen {\O}, Denmark\\
(jslarsen@tac.dk, vedel@tac.dk)\\
\\
$^2$University of California, Lick Observatory,\\
Santa Cruz, CA 95064, USA\\
(uffe@ucolick.org)\\
\\}
\date{}
\maketitle


\section*{Abstract}

A theory for the structure of isothermal, self-gravitating gas spheres in
pressure
equilibrium is developed for softened gravity, assuming an ideal gas equation
of state.
The one-parameter spline softening proposed by
Hernquist \& Katz (1989) is used. We
show
that the addition of this extra scale-parameter implies that the set of
equilibrium
solutions constitute a
one-parameter
family, rather than the one and only one isothermal sphere solution for
Newtonian\\
\\
{\it Submitted to the Astrophysical Journal}\\
\\
gravity.

We develop a number of approximate, analytical or semi-analytical solutions
which apply in various regions of parameter space.

For softened gravity, the structure of isothermal spheres is, in general, very
different from the Newtonian isothermal sphere.
For example, as a corollary,
we demonstrate the perhaps somewhat surprising result that even
in the complete absence
of rotational support it is possible, for {\it any} finite choice of softening
length
$\epsilon$
and temperature $T$, to deposit an arbitrarily large mass of gas in
pressure
equilibrium and with a non-singular density distribution inside of $r_0$
for {\it any} $r_0 > 0$ (neglecting effects of
changes in
the equation of state as well as general relativistic effects).

Furthermore, it is sometimes claimed that the size of the small
scale, self-gravitating gas structures formed in dissipative Tree-SPH
simulations
is simply of the order
the gravitational softening length. We demonstrate, that this, in general, is
{\it not} correct.

The main purpose of the paper is to
compare the theoretical predictions of our models with the properties of the
small,
massive, quasi-isothermal gas clumps ($r \sim 1 {\rm kpc}$, $M \sim 10^{10}
M_{\odot}$
and $T \simeq 10^4 K$) which form in numerical Tree-SPH simulations of
'passive'
galaxy formation of Milky Way sized galaxies
(i.e. simulations not involving energy and momentum feedback
to the gas from supernova explosions, stellar winds, UV radiation from OB stars
etc.).
We find reasonable agreement, despite the neglect of
effects of rotational support in the models presented in this paper.

We comment on whether the hydrodynamical resolution is sufficient in
our
numerical simulations of galaxy formation involving highly super-sonic,
radiative
shocks and we give a necessary condition, in the form of a simple test,
that the hydrodynamical resolution in any such simulations is sufficient.

Finally we conclude that one should be cautious, when comparing results of
numerical simulations, involving gravitational softening and hydrodynamical
smoothing, with reality.

\newpage

\section{Introduction}
Due to the incredible increase in computing power provided by the computer
industry over the last decade or so it has become possible to attempt to model,
by 3-D numerical simulations, the formation and evolution of galaxies using
various combinations of gravitational and hydrodynamical codes (e.g. Evrard
1988,
Hernquist \& Katz 1989, HK89 in the following).
Since the physical problem involves very large density
contrasts fully Lagrangian codes, like Tree-SPH (HK89) which is completely
particle based, are particularly well suited for this problem. The Tree-SPH
code
calculates gravitational interactions using a hierarchical tree method (Barnes
\& Hut
1986) and the hydrodynamical interactions using the SPH (smoothed particle
hydrodynamics) method (Lucy 1977; Gingold \& Monaghan 1977).

Primarily to suppress effects of two-body gravitational interactions in such
simulations
the gravitational field of a gas or dark matter particles is softened,
typically
by
using a Plummer softening kernel (Evrard 1988) or a spline softening kernel
(HK89) -
see section 2 for more details about gravitational softening.

In the 'passive' variant of such simulations (i.e. simulations not involving
energy and
momentum feedback
to the interstellar medium due to supernova explosions, stellar winds, UV
radiation from
OB stars etc.) it is generally found that several small, massive gas
clumps ($r \sim 1 {\rm kpc}, m \sim 10^{10} M_{\odot}$ for simulations of Milky
Way
sized galaxies) are formed quite early in the simulations and survive for
several
Gyrs. The high density gas is nearly isothermal with a temperature $T \simeq
10^{4} K$. This is due to the form of the radiative cooling function used in
the
simulations.  The primordial gas cools quite effectively at temperatures $T
\sim
10^{4}-10^{6}
K$, mainly through collisional excitation of H and ${\rm He}^+$ by free
electrons,
but the cooling function is effectively truncated
below  $T \simeq 10^{4} K$ where the density of free electrons rapidly tends
to zero as the gas recombines.
For low gas densities the radiative
cooling can be suppressed, at the relevant temperatures $T \sim 10^{4}-10^{6}
K$, by
up to two orders of magnitude by inclusion of effects of a background,
ionizing,
UVX radiation field, but at the high densities characteristic
of the small, massive gas clumps the cooling function is essentially unchanged
(Efstathiou 1992, Vedel, Hellsten \& Sommer-Larsen 1994, VHSL94 in the
following).

At first glance it seems somewhat surprising that such small, massive,
isothermal
gas clumps can be in quasi-equilibrium at a temperature of $T \simeq 10^{4} K$,
since for Newtonian gravity the mass inside of $r$ of a stationary, isothermal,
self-gravitating
gas sphere is
\begin{equation}
m(< r) \la ~\frac{2 k_B T }{\mu m_{p} G} ~r = 3.2 ~10^{7}  ~\frac{T_4}
{\mu_{1.2}} ~r  ~~M_{\odot} \;\;,
\end{equation}
where $k_B$ is Boltzmann's constant, $T$ the temperature,
$\mu$ is the mean molecular weight,
$m_p$ is the proton mass, $G$ is the
gravitational constant, $T_{4}$ is the temperature in units of $10^{4} K$,
$\mu_{1.2}$
the mean molecular weight in units of 1.2 and $r$ is
in units of ${\rm kpc}$ in the last expression.
Note, though, that, as shown by Ebert (1955) and Bonner (1956), the Newtonian
isothermal sphere is only stable for
\begin{equation}
r < ~1.72 ~\sqrt{\frac{k_B T }{\mu m_{p} G \rho_0}}~ = 2.03 ~r_{KING} = 1.21
~T_4^{0.5}\mu_{1.2}^{-0.5} ~n_0^{-0.5}  ~~{\rm kpc} \;\;,
\end{equation}
where $\rho_0$ is the central density, $r_{KING}$ is given by equation (13) and
$n_0$ is the central number density of hydrogen in units of $cm^{-3}$.

The dense, massive gas clumps are generally supported by rotation to some
extent,
but it is obviously of relevance to investigate whether such dense, massive gas
clumps
could be in equilibrium even in the absence of any rotational support at all,
as a
consequence of the gravitational softening. In the following it will be
demonstrated
that indeed they can.

In section 2 gravitational softening is briefly discussed.
In section 3 pressure equilibrium solutions for isothermal, self-gravitating
gas spheres will
be derived for softened gravity. In section 4 the theoretical predictions  will
be discussed in
relation to the results of numerical Tree-SPH simulations of galaxy formation.
Section 5 constitutes the conclusion and finally in Appendix A the softened
gravitational
potential and gravitational field of an infinitely thin,  spherical shell is
determined.

\section{Gravitational softening}

One popular way of incorporating gravitational softening in
numerical,
particle based simulations is to replace the Newtonian potential, $\varphi(r) =
- G m/r$,
of a point mass of mass
$m$ with the spherically symmetric Plummer potential
$\varphi_P(r) = - G m/\sqrt{r^2 + b^2}$,
where $b$ is the Plummer scale-parameter. From Poissons equation it follows
that the
point mass is smeared out into a density distribution of total mass $m$
\begin{equation}
\rho_P (r) = \frac{3 m}{4 \pi b^3} (1 + \frac{r^2}{b^2})^{-5/2} ~~,
\end{equation}
which is called the Plummer sphere - see e.g. Binney \& Tremaine (1987).
Note that, for $b > 0$,  $\rho_P (r) > 0$ everywhere. This is to some extent a
disadvantage, at least if the gravitational interactions are calculated with a
Tree
code - see the discussion in HK89.

HK89 proposed, inspired by the work of Gingold \& Monaghan (1977), to
soften the gravitational field of a point mass by using the same spline kernel
as was used in their implementation of SPH. This spherically symmetric spline
(softening) kernel,
originally introduced by Monaghan \& Lattanzio (1985), has the advantage
that it has compact support at $r = 2 \epsilon$, where the scale-parameter
$\epsilon$ is denoted the softening length. The point mass of mass $m$ is hence
smeared out into a density distribution of total mass $m$
\begin{eqnarray}
                               &   1 - 1.5 u^2 + 0.75 u^3 ~~~~~~~~~
~~~~~~~~ & ~0 \le u < 1 \nonumber \\
\rho_{spline}(r) =  ~~~~~~~~~~\frac{m}{\pi \epsilon^3} ~\Bigg\{ & 0.25 ~(2 -
u)^3 ~~~~~~~~~~~~~~~~
~~~~~~~& ~1 \le u < 2  \nonumber \\
                                &  0~~~~~~~~~~~~~~~~~~~~~~~
~~~~~~~~~~~~~~~ & ~~u \ge 2 ~~,\nonumber \\
                                &   &
\end{eqnarray}
where $u = r/\epsilon$. Note that $\rho_{spline} (r) = 0$ for $u \ge 2$, so
outside of
$r = 2 \epsilon$ the gravitational field and potential are exactly the same as
for Newtonian
gravity.

In our Tree-SPH simulations we use the HK89 spline softening for the reasons
given above. We shall consequently adopt this type of gravitational softening
in the following theoretical considerations also.

\section{Equilibrium solutions for isothermal, self-gravitating
gas spheres for softened gravity}

We shall restrict the following analysis to spherically symmetric
systems only:

The force equation for a stationary, isothermal, self-gravitating
sphere gives
\begin{equation}
\frac{d\vec{u}}{dt} = -\frac{\nabla P}{\rho} - \nabla \varphi = \vec{0}
\end{equation}
where $\vec{u}$ is the gas velocity, $P$ the gas pressure, $\rho$ the
gas density and $\varphi$ the gravitational potential.

Throughout this paper we assume that the hydrodynamics are described by
the equations of ordinary gas physics.

In SPH the gas density (and effectively also all other hydrodynamical variables
like gas pressure, energy, entropy etc.) is estimated on the basis of a
smoothing kernel, which,
for example, can be of the form given in equation (4) - see e.g. HK89 and
references
therein for further details. The characteristic smoothing scale is denoted the
smoothing
length $h$. When comparing the theoretical models, developed in the following,
with the small, massive, quasi-isothermal gas clumps, formed in our Tree-SPH
simulations, we assume, in this paper, that the theoretical predictions are
directly applicable.
This assumption is valid here, since $h$ is always much smaller than the
characteristic size of clumps.
One should note, though, that this assumption will not always be valid.

Assuming an ideal gas equation of state
\begin{equation}
P = N k_B T
\end{equation}
where $N$ is the gas number density,
equations (5) and (6) yield for Newtonian gravity
\begin{equation}
\tilde{T} \frac{d \ln \rho}{d r} = -\frac{d \varphi}{d r} =
- \frac{G M(r)}{r^2} \;\;,
\end{equation}
where
\begin{equation}
\tilde{T} = \frac{k_B T}{\mu m_p} = \gamma^{-1} c_s^2 ~~,
\end{equation}
is a constant, since T in the following is assumed to be constant,
and where $\gamma$ is the adiabatic index, $c_s$ the sound speed and
\begin{equation}
M(r) = 4 \pi \int_{0}^{r} \rho(r') r'^2 d r' ~~.
\end{equation}

In equation (7) we use Newton's first and second theorems,
that the gravitational acceleration $g(r)$
depends only on the {\it amount} of mass {\it inside} of $r$ .

In the case of softened gravity the situation is very different: for a
given softening length $\epsilon$ the softened gravitational acceleration
$g_{\epsilon}(r)$ depends on the {\it density distribution} $\rho(r')$
for $r' \in [0 , r + 2 \epsilon]$, so in particular also on the density
distribution out to two softening lengths {\it outside} of $r$.

As will be shown in the following, the
change of the structure of isothermal spheres for softened gravity relative
to the Newtonian case is, in general, quite dramatic.

For Newtonian gravity all solutions to equation (7) can be rescaled
such that equation (7) effectively has one and only one solution in terms of
the normalized density and radius
\begin{equation}
\tilde{\rho}_{KING}(\tilde{r}) = \frac{\rho(r)}{\rho_0} ~~,
\end{equation}
where
\begin{equation}
\rho_0 = \rho(r=0)
\end{equation}
and
\begin{equation}
\tilde{r} = \frac{r}{r_{KING}}
\end{equation}
and where
\begin{equation}
r_{KING} = \sqrt{\frac{9 \tilde{T}}{4 \pi G \rho_0}} = 0.595 ~T_4^{0.5}
~\mu_{1.2}^{-0.5}
~n_0^{-0.5} \;\; {\rm kpc}
\end{equation}
usually is denoted the King radius (e.g. Binney \& Tremaine 1987).

In the case of softened gravity the gravitational field depends on one more
parameter than for Newtonian gravity: $\epsilon$, the softening length. One
would consequently expect that all solutions to the force equation (7) can be
rescaled in terms of a {\it one-parameter family} of solutions and, as we show
in the following, this indeed is the case.

\subsection{Scaling properties of the solutions}

Let $(\epsilon, \tilde{T}, \rho_o)$ be given and let $\rho(r)$ be a solution to
\begin{equation}
\tilde{T} \frac{d \ln \rho}{d r} = - g_\epsilon (r) ~~,
\end{equation}
where
\begin{equation}
g_\epsilon (r) = 4 \pi \int_{0}^{r + 2 \epsilon} \tilde{g}_\epsilon (r,r')
\rho(r') r'^2 d r'
\end{equation}
and where $\tilde{g}_\epsilon (r,r')$ is the softened gravitational
acceleration at $r$
due to
a unit mass, infinitely thin, spherical shell located at $r'$. In Appendix A
an
expression for $\tilde{g}_\epsilon (r,r')$ is derived.

It is straightforward to show that
\begin{equation}
\tilde{g}_{\epsilon'} (\alpha r, \alpha r') = \alpha^{-2} \tilde{g}_\epsilon
(r, r') ~~,
\end{equation}
where
\begin{equation}
\alpha = \frac{\epsilon'}{\epsilon} \;\;.
\end{equation}
Using this it is easy to show that, for parameters $(\epsilon' = \alpha
\epsilon,
\tilde{T}, \rho_0 ' = \alpha^{-2} \rho_0)$,
\begin{equation}
\rho'(r) = \alpha^{-2} \rho(\alpha^{-1} r)
\end{equation}
is a solution to equation (14). In other words if we know all solutions to
equation (14)
for a given softening length $\epsilon$, then all solutions, for {\it any}
softening length
$\epsilon' = \alpha \epsilon$, can be obtained by a mere scaling.
Now, let again $(\epsilon, \tilde{T}, \rho_o)$ be given and let $\rho(r)$ be a
solution to
equation (14). It is straightforward to show that, for parameters $(\epsilon,
\tilde{T}' = \beta \tilde{T}, \rho_0 ' = \beta\rho_0)$,
\begin{equation}
\rho'(r) = \beta \rho(r)
\end{equation}
is a solution to equation (14). Again, if we know all solutions to equation
(14) for a given
value of $\tilde{T}$, then all solutions for {\it any}
$\tilde{T}' = \beta \tilde{T} $ can be obtained by a mere scaling.

Consequently, in general, the solutions to equation (14) can be expressed as a
one-parameter
family in terms of a normalized density
 \begin{equation}
\tilde{\rho}(r'; \rho_0' ) = \frac{\rho}{\rho_0} ~~,
\end{equation}
where
\begin{equation}
r' = \alpha^{-1} r
\end{equation}
and
\begin{equation}
\rho_0' = \alpha^{2} \beta^{-1} \rho_{0} \;\;.
\end{equation}

\subsection{The linear approximation and self-similarity}

Fig. 1 shows the softened gravitational field, $g_\epsilon (r)$,
for shells located at $u_{shell} = 0.0, 0.10, 0.30, 0.50$ and  1.00, where
$u_{shell}$ = $r_{shell}/\epsilon$. For $u_{shell} \la 0.5$, the gravitational
acceleration
at $r \la \epsilon/2$ due to such a shell is approximately linear in $r$:
\begin{equation}
g_\epsilon (r) \simeq -k r M + O(r^3) , ~~{\rm for} ~~r \la \epsilon/2 ~~{\rm
and}
~~r_{shell} \la \epsilon/2 ~~,
\end{equation}
where it follows from the Appendix of HK89 that the constant $k$
is given by
\begin{equation}
k = \frac{4 G}{3 \epsilon^3 } \;\; .
\end{equation}
Equation (14) can then be solved analytically assuming that the scale of the
system is
less than or comparable to $\epsilon/2$:
\begin{equation}
\rho (r) = \rho_0 \exp(-a r^2 ) ~, ~~{\rm for} ~~ r \la \epsilon/2 ~~,
\end{equation}
where
\begin{equation}
a = \frac{k M}{2 \tilde{T}} \;\; .
\end{equation}
If, at $r \simeq \epsilon/2$, $\rho(r) << \rho_0$,
then from the consistency relation for the total mass
\begin{equation}
M \simeq M_\infty = 4 \pi \rho_0 \int_{0}^{\infty} \exp(-a r'^2) r'^2 dr' ~~,
\end{equation}
it follows that
\begin{equation}
M \simeq 2 \pi \rho_0 \Gamma(\frac{3}{2}) a^{-3/2}  \;\; .
\end{equation}
$M$ can then be obtained as
\begin{equation}
M \simeq M_{TAC} = (\frac{3 \pi \tilde{T} \epsilon^3 }{2 G})^{3/5} \rho_0^{2/5}
=
3.88 ~10^8 ~T_4^{0.6} ~\mu_{1.2}^{-0.6} ~n_0^{0.4} ~\epsilon_3^{1.8}
  ~~M_\odot ~~,
\end{equation}
where $n$ is the number density of hydrogen in units of $cm^{-3}$,
$n_0 = n(r=0)$ and $\epsilon_3$ is the softening
length in units of 3 kpc.

From equations (24), (26) and (29) it follows that
\begin{equation}
a = (\frac{2}{3})^{2/5} ~\pi^{3/5} ~G^{2/5} ~\tilde{T}^{-2/5} ~\epsilon^{-6/5}
~\rho_0^{2/5}  \;\; .
\end{equation}
We now introduce the characteristic length scale $r_{TAC}$:
\begin{equation}
r_{TAC} = 3 ~(2 a)^{-1/2} = ~2.74 ~T_4^{0.2} ~\mu_{1.2}^{-0.2}
~\epsilon_3^{0.6}
~n_0^{-0.2} ~~{\rm kpc}\;\; .
\end{equation}
Equation (25) can then be rewritten as
\begin{equation}
\rho_{TAC} (r) = \rho_0 \exp(-(\frac{r^2}{2 (r_{TAC}/3)^2}))\;\; ,
\end{equation}
or as
\begin{equation}
\tilde{\rho}_{TAC} (\tilde{r}) = \frac{\rho}{\rho_0} = \exp(-4.5
\tilde{r}^2)\;\; ,
\end{equation}
where $\tilde{r} = r/r_{TAC}$.
Hence the density distribution is gaussian in $r$ with dispersion $r_{TAC}/3$.
So
for
$r_{TAC} << \epsilon$
all solutions can be rescaled in terms
of one and only one solution, as is the case for the Newtonian isothermal
sphere, though
the nature of the two solutions is very different.

It is sometimes claimed that the size of the small
scale, self-gravitating gas structures formed in dissipative Tree-SPH
simulations
is simply of the order
the gravitational softening length. From equation (31) it follows that, in
general, this is
{\it not} correct.

\subsection{The point mass approximation}

From Fig. 1 it can be seen that, for a given softening length $\epsilon$ and a
shell
of mass $m$ located
at $r_{shell} \la \epsilon/2$, a point mass of mass $m$ at $r' = 0$ results in
a softened
gravitational field, which is a quite good approximation to softened
gravitational field of the shell and for $r \ga \epsilon/2$ is a
significantly better approximation, than
the linear approximation discussed in the previous section.

Rewriting equation (14) as
\begin{equation}
\tilde{T} \frac{d \ln \rho}{d r} = -M \frac{d \varphi_{pm}}{d r} \;\;,
\end{equation}
where $\varphi_{pm}$ is the specific potential corresponding to the softened
gravitational
field of a point mass of unit mass, it follows by integration of equation (34)
that
$\rho_{pm}(r)$ can be expressed as
\begin{equation}
\rho_{pm} (r) = \rho_0 \exp(-(\varphi_{pm}(r) - \varphi_{pm}(r=0))
\frac{M}{\tilde{T}}) \;\; .
\end{equation}
In the Appendix of HK89 $\varphi_{pm}(r)$ is expressed as
\begin{equation}
\varphi_{pm}(r) = - G f(r) \;\;;
\end{equation}
the function $f(r)$ is also given as equation (A3) in Appendix A of this paper.

Equation (34)  can hence be rewritten as
\begin{equation}
\tilde{\rho}_{pm} = \frac{\rho}{\rho_0} = \exp((f(r) - f(r=0))
~\frac{GM}{\tilde{T}})  \;\; .
\end{equation}
The consistency equation for the total mass can not be used in this case, since
it can
be shown that
\begin{equation}
M(r) = 4 \pi \rho_0 \int_{0}^{r} \rho_{pm}(r') r'^2 dr' \rightarrow \infty
~{\rm as}
~r \rightarrow \infty ~~.
\end{equation}
However, by setting $M = M_{TAC}$ in equation (37), one obtains, for
systems characterized by
$r_{TAC} \la \epsilon$, a generally much better approximation to the true
solution than
for the linear approximation - see the following subsection.

\subsection{The general solution}

A numerical, iterative algorithm was developed to obtain exact solutions of
equation (14).

The general solutions can conveniently be expressed as
\begin{equation}
\tilde{\rho}(u; u_{TAC}) = \frac{\rho}{\rho_0} ~,
\end{equation}
where $u = r/\epsilon$ and
\begin{equation}
u_{TAC} = \frac{r_{TAC}}{\epsilon} = 0.913 ~T_4^{0.2} ~\mu_{1.2}^{-0.2}
~\epsilon_3^{-0.4}
~n_0^{-0.2} ~~.
\end{equation}

In Fig. 2 the solutions are shown, as functions of $u$, for a large range of
the $u_{TAC}$
parameter.  As $u_{TAC} \rightarrow 0$ the solutions, as functions of $u$,
converge towards a vertical line at $r$ = 0
and as $u_{TAC} \rightarrow \infty$ the solutions, as functions of $u$,
converge
towards a horizontal line at $\tilde{\rho}$ = 1.

As $u_{TAC} \rightarrow 0$, then $r_{TAC}$ becomes the
characteristic linear scale of the isothermal spheres and the solutions
converge as
\begin{equation}
\tilde{\rho}(u;u_{TAC}) \rightarrow \tilde{\rho}_{TAC} (u/u_{TAC}) ~,
~~~~~u_{TAC} \rightarrow 0 ~.
\end{equation}

Conversely, as $u_{TAC} \rightarrow \infty$ $r_{KING}$
becomes the characteristic linear scale of the isothermal
spheres and the solutions converge as
\begin{equation}
\tilde{\rho}(u;u_{TAC}) \rightarrow \tilde{\rho}_{KING} (u/u_{KING}) ~,
~~~~~u_{TAC} \rightarrow \infty ~,
\end{equation}
where $u_{KING} = r_{KING}/\epsilon$ and there is a one-to-one
relation between $u_{KING}$ and $u_{TAC}$:
\begin{equation}
u_{KING} = 0.249 ~u_{TAC}^{2.5} ~~.
\end{equation}

In Fig.3 $\tilde{\rho}/\tilde{\rho}_{TAC}$ is shown, as function of
$r/r_{TAC}$,
for various values
of $u_{TAC}$ and, as expected,
$\tilde{\rho}/\tilde{\rho}_{TAC} \rightarrow 1$ as
$u_{TAC} \rightarrow 0$.

In Fig.4 $\tilde{\rho}/\tilde{\rho}_{KING}$ is shown, as function of
$r/r_{KING}$,
for various values
of $u_{TAC}$ and, likewise as expected,
$\tilde{\rho}/\tilde{\rho}_{KING} \rightarrow 1$ as
$u_{TAC} \rightarrow \infty$.

In Fig. 5 $\tilde{\rho}/\tilde{\rho}_{pm}$ is shown, as a function of
$r/r_{TAC}$, for
various values of $u_{TAC}$ and, as can be seen from the figure,
$\tilde{\rho}/\tilde{\rho}_{pm} \rightarrow 1$ as $u_{TAC} \rightarrow 0$ - as
expected - and it is obvious that the point mass approximation, in general, is
much
better than the linear approximation (Fig. 3), as discussed in subsection 3.3.

Fig. 6 shows $M/M_{TAC}$ as a function of $u = r/\epsilon$ for various
values
of $u_{TAC}$. As can be seen from the figure, $M \simeq M_{TAC}$ for $u_{TAC}
\la 0.5$ whereas for larger values of $u_{TAC}$, $M$ becomes increasingly
larger
than $M_{TAC}$ and increases steadily with $u$.

This behaviour is expected, since
for $u_{TAC} \ga 0.5$ the linear approximation breaks down and for the point
mass
approximation, for the general solution and for the Newtonian isothermal
sphere it can
be shown that $M(u) \rightarrow \infty$ as $u \rightarrow \infty$.

Fig. 7 shows, for $\epsilon$ = 3 {\rm kpc}, a typical value used in current
Tree-SPH
simulations of galaxy formation, and $(T_4/\mu_{1.2})$ = 1, the density
profiles
of the general solution for log($n_0$) = -2, -1, .... , 6.
Similarly Fig. 8 shows, for the same parameters, $M(r)$ in units of $10^8
M_{\odot}$.

As can be seen, the masses are in general much larger, at $r \sim$ 1 kpc, which
is
the typical size of the massive, cold gas clumps formed in our 'passive'
simulations,
than would be expected on the
basis of isothermal sphere solutions for Newtonian gravity - see equation (1).

As a corollary it follows from equations (29) and (31) that, for given
physical parameters $(\epsilon,\tilde{T})$, $r_{TAC} \rightarrow 0$ and
$M_{TAC} \rightarrow \infty$ as $n_0 \rightarrow \infty$. So we obtain the
perhaps
somewhat surprising result that for {\it any}
$\epsilon > 0$ and {\it any} positive value of
$(T_4/\mu_{1.2})$ - like $\simeq 1$ for typical small, massive, cold clumps
formed in
Tree-SPH simulations of 'passive' galaxy
formation - it is possible to deposit an arbitrarily large mass of gas
in pressure equilibrium and with a non-singular density distribution inside of
$r_0$,
for {\it any} $r_0 > 0$ (neglecting effects of changes in the equation
of state as well as general relativistic effects).

\section{Comparing with the results of numerical Tree-SPH simulations}

To compare the theoretical predictions obtained in the previous section with
what is actually
found in numerical Tree-SPH simulations of 'passive' galaxy formation, we
performed
a series
of simulations, starting from vacuum boundary, top-hat initial conditions,
similar to
those described in VHSL94. These simulations are not very realistic with
regards to galaxy formation, but this is of no consequence for the present
purpose.

We carried out four simulations. The first three only differed by having
$\epsilon_{SPH}$
= 1.5, 3.0 and 6.0 kpc respectively. The fourth was a higher resolution
simulation - it
was identical to the second simulation except that 8 times more gas particles
(each with
one eighth of the original gas particle mass) were used (but note that {\it no}
extra phases
were added
to the initial conditions). The purpose of this simulation was to check whether
the
hydrodynamical resolution was sufficient in our simulations.
Other details about the simulations
are given in Table 1.

In each of the four simulations we selected the three most massive, cold gas
clumps
($(T_4/\mu_{1.2}) \simeq 1)$ at time $t = 3.2$ Gyr. At this time all the cold
gas
clumps
were self-gravitating. They typically form at $t \la 1$ Gyr in 'mini' dark
matter halos, but
the dark matter is stripped off during the violent relaxation
phase at first recollapse of the proto-galaxy at $t \simeq 2.2$ Gyr.

All the cold gas clumps had, at $t = 3.2$ Gyr, a radial extent of less than or
order
$\epsilon$. The
cumulative mass distribution, at $t = 3.2$ Gyr, for the largest, cold gas clump
in
simulations \#1-3, is shown in Fig. 9 and the cumulative mass distribution, at
$t = 3.2$
Gyr, for simulations \#2 and \#4 (low versus high hydrodynamical resolution),
is shown
in Fig. 10 for all three clumps.

For each cold gas clump we measured the mass $M_{obs}$ and the central density
$n_0$ and determined the
mass $M_{teo}$, which would be expected for an isothermal sphere of this
central
density and parameters $\epsilon_3$ and $(T_4/\mu_{1.2})$ as in the simulations
(all gas clumps were in the $u_{TAC} \la 1$ regime). The ratio
$M_{teo}/M_{obs}$
for the individual clumps is plotted in Fig. 11. As can be seen from the
figure,
$M_{teo}/M_{obs}$ is generally well below unity, indicating that effects of
rotational
support, which are not included in our theoretical models, is
of importance. For all the clumps $M_{teo}/M_{obs}$ is observed to increase as
$\epsilon$
increases.  This is to be expected since, for clumps of similar mass and
specific angular
momentum, the ratio $q$ of pressure force to total gravitational
force scales approximately as $\epsilon^{3/2}$ when $q$ is well below unity
and $u_{TAC} \la 1$.
Furthermore, as the clump masses decrease, $M_{teo}/M_{obs}$ is observed to
increase for a given $\epsilon$. This is also to be expected, though on rather
more
complicated grounds: Theoretically one would expect that the average
dimensionless
spin-parameter $<\lambda>$ depends only weakly on the magnitude
of the initial density fluctuations later collapsing and forming the 'mini'
dark matter
halos  in which the gas clumps subsequently form dissipatively (e.g. Barnes \&
Efstathiou 1987). If one furthermore assumes
that the mass of a gas clump depends on the circular speed $v_c$ of its 'mini'
dark matter halo like  $M \propto v_c^{\alpha}, ~\alpha \simeq 3$, from e.g.
the Tully-Fisher relation, then one can show that, for a given softening
length, $q$ scales approximately as $M^{-2}$ for $q$ well below unity and
$u_{TAC} \la 1$.

For the smallest clump in simulations \#1-4 the effect of pressure forces
appears to be
comparable to the effect of rotational support. In fact, for softened gravity,
the relationship
between the physical structure of self-gravitating, rotationally
supported systems and the gravitational
softening length $\epsilon$ is quite similar to what is found in this paper for
isothermal
spheres, as shown in Sommer-Larsen, Vedel \& Hellsten (1997).

Finally a note on hydrodynamical resolution in Tree-SPH simulations:
From
Figs. 10 and 11 it seems - comparing simulations \#2 and \#4 - that the
hydrodynamical resolution is sufficient. At later times, when the clumps have
merged and super-sonic, radiative  shocks have occurred, the results of
simulations \#2 and \#4 are still very similar - as an example the surface
density
in the resulting gas disks, at $t = 5.2$ Gyr, is shown in Fig. 12 for
simulations
\#2 and
\#4. This is quite reassuring, because, whereas in nonradiative shocks the
density
contrast always is less than the Rankine-Hugoniot limit of
$(\frac{\gamma+1}{\gamma-1})$,
where $\gamma$ is the adiabatic index,  significantly larger
density contrasts most probably occur in the radiative shocks and this
could potentially cause spurious effects, because of
effects of limited shock resolution due to the smoothing, inherent in SPH.

We propose that one should always test a SPH simulation in the manner
described above for simulations \#2 and \#4. Such a simple test constitutes a
necessary (but not, in general, sufficient) condition that the hydrodynamical
resolution is sufficiently high to adequately resolve super-sonic, radiative
shocks.

\section{Conclusion}

We have developed a theory for the structure of isothermal, self-gravitating
gas
spheres in pressure equilibrium for softened gravity, based on the spline
softening
kernel proposed by Hernquist \& Katz (1989).

Because the gravitational force depends
on an extra scale-parameter, the softening length $\epsilon$, relative to
Newtonian
gravity, the solutions constitute a one-parameter family, rather than the one
and only
one isothermal sphere solution for Newtonian gravity.

For softened gravity the structure of isothermal spheres is, in general, very
different
from the Newtonian, isothermal sphere.

For example we find, as a corollary,
the perhaps somewhat surprising result, that,
for {\it any} finite softening length $\epsilon$ and temperature $T$,
it is possible to deposit an arbitrarily large
mass of gas, in pressure equilibrium and with a non-singular
density distribution, inside of $r_0$, for {\it any}
$r_0 > 0$
(neglecting effects of changes in the equation of state and general
relativistic effects).

Furthermore, it is sometimes claimed that the size of the small
scale, self-gravitating gas structures formed in dissipative Tree-SPH
simulations
is simply of the order
the gravitational softening length. We demonstrate, that this, in general, is
{\it not} correct.

The main purpose of the paper is to
compare the theoretical predictions of our models with the properties of the
small, massive,
quasi-isothermal gas clumps formed in numerical Tree-SPH simulations of
'passive' galaxy formation of Milky Way sized galaxies
(i.e. simulations not involving energy and momentum
feedback to the gas from supernova explosions, stellar winds, UV radiation from
OB stars etc.). We find reasonable agreement, despite the neglect
of
effects of rotational support in our theoretical models.

We would expect that if the gravitational softening was based on a Plummer
kernel
(which is also a one scale-parameter type of softening), rather than the spline
kernel,
the resulting isothermal spheres would be qualitatively similar to those
described
in this paper. We encourage any Plummer softening 'fan' to check this
quantitatively
by going through a similar exercise as described in this paper for the spline
softening.

We have not discussed the stability of the isothermal sphere equilibrium
solutions
obtained in this work - we shall return to this issue in a forthcoming paper
using
an approach similar to that outlined by Ebert (1955) and Bonner (1956).

We comment on whether the hydrodynamical resolution is sufficient in
our
numerical simulations of galaxy formation involving highly super-sonic,
radiative
shocks and we give a necessary condition, in the form of a simple test,
that the hydrodynamical resolution in any such simulations is sufficient.

The results obtained in this paper,
and in Sommer-Larsen, Vedel \& Hellsten (1997), seem to indicate that one
should
be cautious when comparing what is observed in the real Universe with
results of numerical Tree-SPH simulations or any
other numerical, gravitational-hydrodynamical simulations, where the
calculation of the gravitational interactions, locally or globally, is particle
based and
the gravitational field of the individual particles is softened.

To summarize: Without proper testing/understanding of the effects of
gravitational
softening and hydrodynamical smoothing in numerical
gravitational-hydrodynamical
simulations of various physical problems, the degree of reality represented by
such
simulations may be quite difficult to assess.

\section*{Acknowledgements}

We have benefitted considerably from discussions with Cedric Lacey and
Draza Markovi\'{c}.
This work was supported by Danmarks Grundforskningsfond through its support
for an establishment of the Theoretical Astrophysics Center.
UH acknowledges support by a postdoctoral research grant from the
Danish Natural Science Research Council.
\newpage

\begin{table}[p]
\caption{Tree-SPH simulations \#1-4.}
\begin{center}
\begin{tabular}{lrrrr}
\hline\hline
Simulation  & $\epsilon_{SPH}$ & $\epsilon_{DM}$ & N$_{SPH}$ & N$_{DM}$ \\
\hline
                         & [kpc]                            & [kpc]
                           &                          &          \\
\hline
\#1                 & 1.5                                 & 10.0
                            &  2048              & 2048 \\
\#2                 & 3.0                                 & 10.0
                            &  2048              & 2048 \\
\#3                 & 6.0                                 & 10.0
                            &  2048              & 2048 \\
\#4                 & 3.0                                 & 10.0
                            & 16384              & 2048 \\
\hline \hline
\end{tabular}
\end{center}
\end{table}

\newpage
\section*{References}
\begin{trivlist}
\item[] Barnes, J., Efstathiou, G., 1987, ApJ, 319, 575
\item[] Barnes, J., Hut, P., 1986, Nature, 324, 446
\item[] Binney, J., Tremaine, S., 1987, Galactic Dynamics. Princeton Univ.
Press,
\item[] \hspace{1.0cm} Princeton
\item[] Bonner, W.B., 1956, MNRAS, 116, 351
\item[] Ebert, R., 1955, Zs.Ap., 37, 217
\item[] Efstathiou, G., 1992, MNRAS, 43p
\item[] Evrard, A.E., 1988, MNRAS, 235, 911
\item[] Gingold, R.A., Monaghan, J.J., 1977, MNRAS, 181, 375
\item[] Hernquist, L., Katz, N., 1989, ApJS, 70, 419 (HK89)
\item[] Lucy, L., 1977, AJ, 82, 1013
\item[] Monaghan, J.J., Lattanzio, J.C., 1985, A\&A, 149, 135
\item[] Sommer-Larsen, J., Vedel, H., Hellsten, U., 1997, in preparation
\item[] Vedel, H., Hellsten, U., Sommer-Larsen, J., 1994, MNRAS, 271, 743
(VHSL94)
\end{trivlist}

\newpage

\appendix
\section*{Appendix A: The softened gravitational potential and gravitational
field of
an infinitely thin, spherical shell}

It is fairly easy to show that the potential, at radial coordinate $r$, of an
infinitely thin,
spherical shell with radius $R$ and unit mass is given by
\begin{eqnarray}
                                        & \frac{1}{2 R r} \int_{R-r}^{R+r}
\varphi_{pm}(r') r' dr' ~ , & ~~r < R
\nonumber \\
\varphi_{shell}(r,R)  =  \Bigg\{ &
                                  &    \\
                                         & \frac{1}{2 R r} \int_{r-R}^{r+R}
\varphi_{pm}(r') r' dr' ~ , & ~~r \ge R
\nonumber
\end{eqnarray}
where $\varphi_{pm}$ is the softened potential of a point mass of unit mass.
In the Appendix of HK89 $\varphi_{pm}$ is expressed as
\begin{equation}
\varphi_{pm}(r) = -G f(r)
\;\; ,
\end{equation}
where
\begin{eqnarray}
                               & -[2 u^2/3 - 3 u^4/10 +
u^5/10]/\epsilon+1.4/\epsilon ~~~~~~~~~~~~
~~~~~~~~ & ~0 \le u < 1 \nonumber \\
f(r) =  \Bigg\{ & -1/(15 r) - [4 u^2/3 -u^3 + 3 u^4 /10 - u^5 / 30]/\epsilon +
1.6/\epsilon & \
1 \le u < 2  \nonumber \\
                                &
1/r~~~~~~~~~~~~~~~~~~~~~~~~~~~~~~~~~~~~~~~~~~~~~~~~
~~~~~~~~~~~~~~~ & ~~u \ge 2 \nonumber \\
                                &   &
\end{eqnarray}

The gravitational acceleration, at $r$, from the unit mass shell at $R$, is
then
given by


\begin{eqnarray}
\tilde{g}_{\epsilon}(r,R) = & - ~d \varphi_{shell}/d r =
~~~~~~~~~~~~~~~~~~~~~~~~~~~~~
~~~~~~~~~& \\
\nonumber \\
					& \frac{1}{2 R r^2} \int_{R-r}^{R+r}
\varphi_{pm}(r') r' dr'
  ~~~~~~~~~~~~~~~~ & \nonumber \\
                                        & -\frac{1}{2 R} (\varphi_{pm}(R+r) -
\varphi_{pm}(R-r)) ~~ &
\nonumber \\
                                        & - \frac{1}{2 r} (\varphi_{pm}(R+r)
+ \varphi_{pm}(R-r)) ~~&
~~~~~~~r < R ~, \nonumber \\
{\rm and}~~~~~~ & & \nonumber \\
					& \frac{1}{2 R r^2} \int_{r-R}^{r+R}
\varphi_{pm}(r') r' dr'
  ~~~~~~~~~~~~~~~~ & \nonumber \\
                                        & -\frac{1}{2 R} (\varphi_{pm}(r+R) -
\varphi_{pm}(r-R)) ~~ &
\nonumber \\
                                        & - \frac{1}{2 r} (\varphi_{pm}(r+R)
+ \varphi_{pm}(r+R)) ~~&
~~~~~~~r \ge R ~.\nonumber
\end{eqnarray}

It is possible to calculate
$\tilde{g}_{\epsilon}(r,R)$ analytically, but as the result is quite
complicated and  voluminous, we shall omit it here.

From equation (A4) it follows that for a spherical system,
with mass distribution $\rho(r)$,
the gravitational acceleration at $r$ is given by
\begin{equation}
g_{\epsilon}(r) = 4 \pi  \int_0^{r + 2 \epsilon} \tilde{g}_{\epsilon}(r,r')
\rho(r') r'^2 dr'  \;\; .
\end{equation}

\newpage
\section*{Figure captions}
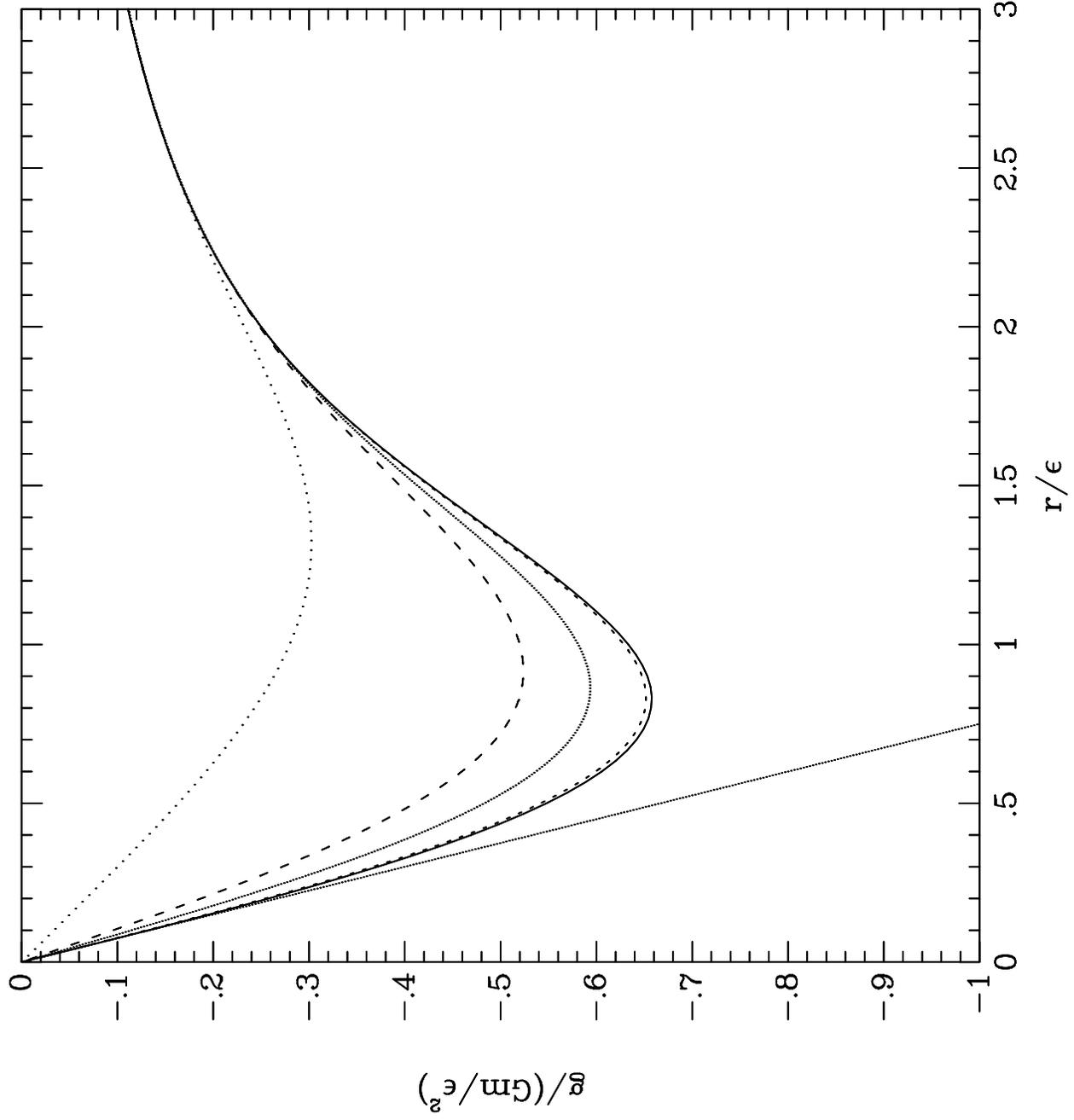
Figure 1:  Gravitational acceleration due to infinitely thin, spherical shells
of mass
$m$ located at $u_{shell}$ = $r_{shell}/\epsilon$
= 0.00, 0.10, 0.33, 0.50 and 1.00. Solid line corresponds to $u_{shell}$ = 0.00
and
as $u_{shell}$ increases the curves move monotonically upwards.
The linear approximation is shown as the straight dotted line.\\
\\
Figure 2:  The general solutions for isothermal spheres for softened gravity.
As $u_{TAC}$ increases the solutions become monotonically more horizontal.\\
\\
Figure 3:  $\rho/\rho_{TAC}$ as a function of $r/r_{TAC}$. The solid,
horizontal,
straight line corresponds to $u_{TAC}$ = 0.00. As $u_{TAC}$ increases the
curves
move monotonically towards the other solid line, which represents $u_{TAC}$ =
$\infty$.\\
\\
Figure 4:  $\rho/\rho_{KING}$ as a function of $r/r_{KING}$. As $u_{TAC}$
increases the curves converge towards the solid, horizontal,
straight line,
which corresponds to $u_{TAC}$ = $\infty$. The other solid line corresponds to
$u_{TAC}$ = 0.00.\\
\\
Figure 5:  $\rho/\rho_{pm}$ as a function of $r/r_{TAC}$. As $u_{TAC}$
decreases the
curves converge towards the solid, horizontal, straight line,
which
corresponds to $u_{TAC}$ = 0.00\\
\\
Figure 6:  $M/M_{TAC}$ as a function of $r/\epsilon$. As $u_{TAC}$ increases
the curves start out monotonically less steep.\\
\\
Figure 7:  Density distributions for isothermal spheres, of various central
hydrogen number density $n_0$, for $T = 10^4$ K and gravitational softening
length
$\epsilon = 3.0$ kpc.\\
\\
Figure 8:  Cumulative mass distribution for isothermal spheres, of various
central hydrogen number density $n_0$, for $T = 10^4$ K and gravitational
softening length
$\epsilon = 3.0$ kpc. The lines correspond to the same values of $n_0$ as in
Fig. 7. \\
\\
Figure 9:  Cumulative mass distribution, at $t$ = 3.2 Gyr, for the most
massive,
cold gas clump in simulations \#1-3.\\
\\
Figure 10:  Cumulative mass distribution, at $t$ = 3.2 Gyr, for the three most
massive,
cold gas clumps in simulations \#2 and \#4.\\
\\
Figure 11:  The ratio of the theoretical to observed mass, at $t$ = 3.2 Gyr,
for the three
most massive, cold gas clumps in simulations \#1-4. For more details - see
text.\\
\\
Figure 12:  The gas surface density distributions, at $t$ = 5.2 Gyr, for the
disks
formed in simulations \#2 and \#4.\\

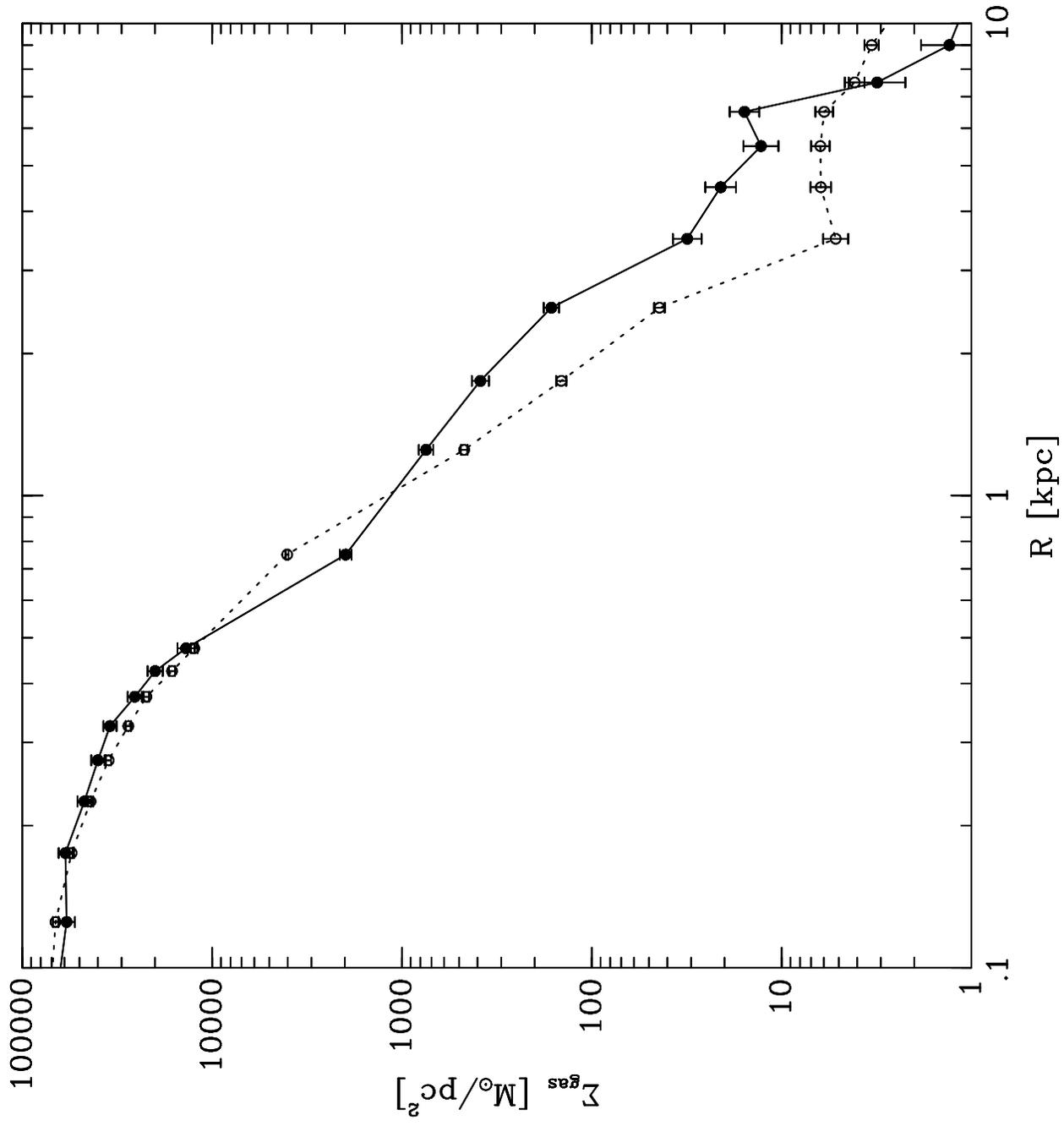
\end{document}